\documentclass[letter,twocolumn]{article}
\usepackage{authblk}
\usepackage{amsmath}
\usepackage{txfonts}
\usepackage{comment}
\usepackage{bm}
\usepackage{color}
\usepackage[dvipdfmx]{graphicx}

\title{Cell Motion Alignment as Polarity Memory Effect}

\author{Katsuyoshi Matsushita\thanks{kmatsu@cp.cmc.osaka-u.ac.jp},  Kazuya Horibe, Naoya Kamamoto and Koichi Fujimoto}
\affil{Department of Biological Sciences, Osaka University, Toyonaka, Osaka, Japan \\} 

\begin{document}
\maketitle

\begin{abstract}
The clarification of the motion alignment mechanism in collective cell migration is an important issue commonly in physics and biology. 
In analogy with the self-propelled disk, the polarity memory effect of eukaryotic cell is a fundamental candidate for this alignment mechanism.
In the present paper, we theoretically examine the polarity memory effect for the motion alignment of cells on the basis of the cellular Potts model. 
We show that the polarity memory effect can align motion of cells. 
We also find that the polarity memory effect emerges for the persistent length of cell trajectories longer than average cell-cell distance.
\end{abstract}

Motion alignment plays various roles widely in self-propelled systems including migrating cells\cite{Entschladen:2009}, moving organisms \cite{Vicsek:2012}, molecular motors\cite{Schaller:2010},  self-propelled droplets \cite{Seeman:2016} and swarming robots \cite{Brambilla:2013}. 
In particular, the alignment of migrating cells is indispensable for cell organizing in organogenesis, wound healing and immune response \cite{Weijer:2009,Friedl:2009,Rorth:2009}.
In these processes, migrating cells exhibit collective behavior commonly observed in self-propelled systems \cite{Romanczuk:2012, Marchetti:2013, Doostmohammadi:2018}, including various patterns \cite{Wakita:2015, Kawaguchi:2017}, active turbulence \cite{Wensink:2012}, traveling wave excitation \cite{Kuwayama:2013}. 
For the understanding of these behavior,  an important issue is to clarify the alignment mechanism as their underlying basis. 

The alignment mechanisms of other self-propelled systems may give hints for this clarification.
In many self-propelled systems including bird flocking \cite{Reynolds:1987, Vicsek:1995}, the direct aligning-interaction through visual contact is supposed and is expressed as an interaction between vector degrees of freedom, which are so-called polarity.
Unfortunately, this direct aligning-interaction 
is not expected in migrating cells lacking visual contact.
As another candidate, the peculiar alignment of self-propelled disks or deformable particles is considerable\cite{Weber:2013, Lam:2015, Yamanaka:2014, Ohta:2017} because it requires only an indirect interaction of polarities through excluded volume. 
In these cases, the direct aligning-interaction is not necessary at least for this alignment. Especially for the disks, the polarity memory recording the past motion aligns the subsequent motion through collisions even with rotation-symmetric excluded volume \cite{Hanke:2013, Hiraoka:2016, Hiraoka:2017, Martin-Gomez:2018}.

Since the polarity memory recording cell motion is well known for eukaryotic cells\cite{Petrie:2009, Trepat:2018},
the polarity memory effect is expected to work in the collective cell migration of {\it Dictyostelium Discoideum} (Dicty)\cite{Takagi:2008,Li:2011}, keratocytes\cite{Szabo:2006, Maree:2006} and neural crest cell\cite{Szabo:2016} . 
In the studies of these collective migration \cite{Mehes:2013,Haeger:2015}, since the alignment was attributed only to the chemataxis\cite{Bagorda:2008, Akiyama:2017}, the contribution of polarity memory has been overlooked. 
In addition, even when the chemotaxis is artificially inhibited \cite{Kuwayama:1993, Kuwayama:2013}, other effects including intercellular adhesion \cite{Ilina:2009, Matsushita:2018}, shape anisotropy\cite{Merks:2006, Starruss:2007, Hirashima:2017} or contact inhibition\cite{Mayor:2010, Schnyder:2017}/activation\cite{Fujimori:2019, Hiraiwa:2019} of locomotion have the polarity memory effect be experimentally invisible. 
Therefore, the theoretical examination to evaluate the polarity memory effect is a powerful method to clarify the alignment mechanism.

In this examination, the spontaneous fluctuation in the shape may give considerable differences between 
eukaryotic cells and self-propelled disks \cite{Blanchard:2010, Munoz:2013}, even when the averaged cell shape is rotation symmetric. 
In particular, since the shape fluctuation due to protrusion \cite{Bignold:1987,Haastert:2011} or blebbing \cite{Fackler:2008,Charras:2008} stochastically propels cells, the fluctuation clearly gives the qualitative difference between cells and self-propelled disks in the propulsion mechanism. 
{\color{red} In fact, the behavior of the cells cannot directly be deduced from the theory of the self-propelled disks, because cells lose propulsion in unfluctuating shape like the disks.}
Therefore, for the theoretical examination, the theory, only by itself, is insufficient. 
At least, it should be combined with the cellular model suitably expressing the propulsion mechanism\cite{Anderson:2007}.

In the present work, we investigate a model for migrating cells with the polarity memory and the interaction only through  rotation-symmetric excluded volume. We perform Monte Carlo simulations based on the Cellular Potts model\cite{Graner:1992, Graner:1993} and thereby confirm that the alignment mechanism due to the polarity memory is effective. Then, we further examine the shape fluctuation effect on the alignment and show that the alignment emerges at the crossover between the persistent length of cell trajectories and the average cell-cell distance when the propulsion is stronger than the shape fluctuation.

Let us consider the two dimensional Cellular Potts model consisting of migrating cells with the polarity memory \cite{Kabla:2012}. This model is defined by the Hamiltonian,
\begin{align}
{\cal H} = {\cal H}_{\rm CC} + {\cal H}_{\rm CE} + {\cal H}_{\rm Vol} + {\cal H}_{\rm Mot} \label{eq:total_hamiltonian}
\end{align}
for given Potts state $\{m(\bm r) \}$ on the square lattice. Here, $\bm r$ represents a site in the square lattice. $m(\bm r)$ takes a number from 0 to the number of cells $N$. $m(\bm r)$ = 0 expresses that the site $\bm r$ is empty. Otherwise, $m(\bm r)$ is the index of cells on the site $\bm r$. For simplicity of examination, we fix  $N$. 

The first and second terms in rhs of Eq.~\eqref{eq:total_hamiltonian},
\begin{align}
{\cal H}_{\rm CC} &= \Gamma\sum_{\left<\bm r \bm r'\right>} \eta_{m(\bm r)m(\bm r')}\eta_{0m(\bm r')}\eta_{m(\bm r)0}, \\
{\cal H}_{\rm CE} &= \Gamma_0\sum_{\left<\bm r \bm r'\right>} \left(\delta_{m(\bm r)0}\eta_{0m(\bm r')} + \eta_{m(\bm r)0}\delta_{m(\bm r')0}\right),
\end{align}
represent interface parts of cell-cell and cell-empty space. $\delta_{mn}$ is Kroneker $\delta$ and $\eta_{mn}$ is defined by $(1 - \delta_{mn})$. $\Gamma$ represents cell-cell interface tension and $\Gamma_0$ cell-empty space interface tension. For $\Gamma$ $>$ $2\Gamma_0$ \cite{Glazier:1993}, cells are suspended. Since the suspended cells interact only with excluded volume, we impose this condition on $\Gamma$ for our purpose.
The summations in rhs of these equations are taken over all the neighboring sets which consist of the nearest and next nearest neighbor sites\cite{Graner:1992}. 

The third term in the rhs of Eq.~\eqref{eq:total_hamiltonian},
\begin{align}
{\cal H}_{\rm Vol} = \kappa \sum_{m=1}^{N}\left(1 - \frac{\sum_{\bm r}\delta_{mm(\bm r)}}{V}\right)^2.
\end{align}
represents the balk part.
This term expresses the situation where the occupation area of cells is maintained to be $V$, as empirically observed. Here, $\kappa$ represents area stiffness.

The forth term in the rhs of Eq.~\eqref{eq:total_hamiltonian},
\begin{align}
{\cal H}_{\rm Mot} = - \varepsilon \sum_{m=1}^{N} \sum_{\bm r} \delta_{mm(\bm r)} \bm p_m \cdot  \bm e_m(\bm r),
\end{align}
represents propulsion \cite{Kabla:2012}. 
Here, $\varepsilon$ is the strength of propulsion, $\bm p_m$ is the unit vector of the polarity and $\bm e_m(\bm r)$ is the unit vector indicating from the center of the $m$-th cell ${\bm R}_{m}$ = $\sum_{\bm r}\bm r\delta_{mm(\bm r)}/\sum_{\bm r}\delta_{mm(\bm r)}$ to a site $\bm r$. 
Notice that the cell takes rotation-symmetric shape even with this term, because this term does not induces tensile stress. $\bm p_m$ obeys \cite{Szabo:2006, Matsushita:2017}
\begin{align}
    \frac{d \bm p_m}{dt} = \frac{1}{a\tau_p\Delta t} \left[ \frac{d {\bm R}_m}{dt} - \left(\bm p_m \cdot \frac{d \bm R_m}{dt}\right) {\bm p}_m \right]. \label{eq:polarity}
\end{align}
Here, $t$ is time, $\tau_p$ is the time scale ratio of $\bm p_m$ change to $\bm R_m$ change and $a$ is the lattice constant. This equation represents the polarity memory during the time of $\tau_p\Delta t$ in the sense of the solution, 
$\bm p_m(t) \sim \int^{t}_{-\infty}dt'd{\bm R}_m(t')/dt'\exp[-(t-t')/\tau_p\Delta t]/a\tau_p\Delta t$\cite{Kabla:2012}.
Here, $\Delta t$ represents the time of single monte carlo steps (mcs) and we set $\Delta t$ = 1 for simplicity.

On the basis of $\cal H$, the shape fluctuation of cells is reproduced by monte carlo simulation \cite{Anderson:2007}. In this simulation,
the state $\{m(\bm r)\}$ is updated to $\{m'(\bm r)\}$ as follows: firstly a site $\bm r$ is randomly selected. Then, the state $m(\bm r')$ at a randomly selected site $\bm r'$ in its neighboring set is copied to the site $\bm r$. This copy is accepted with the 
 Metropolis probability $P(m'(\bm r)| m(\bm r))$ = $\min(1, \exp\left\{-\beta \left[{\cal H}(m(\bm r)) - {\cal H}(m'(\bm r))\right]\right\})$, where $\beta$ is inverse temperature. Otherwise, this copy is rejected. 
We call this procedure a copy trial. 
Mcs conventionally consist of $16L^2$ copy trials \cite{Graner:1992}, where $L$ is the linear size of square system. The series of mcs expresses the shape fluctuation.
{\color{red} For each mcs, $\bm p_m$ is assumed to be a slow variable with $\bm R_m$ and is updated once by the Eular method \cite{Matsushita:2017}. Simultaneously, $\bm R_m$ is also updated and is fixed in mcs.}

In this simulation, we employ the following parameters: 
$L$ is 256 and $V$ is 64.
These parameters are chosen so as to realize the suspended state for the tractable $N$ from 64 to 512. These $N$ corresponds to the area fraction of $\phi$ from 0.11 to 0.85. 
Additionally for the same purpose, we choose surface tensions $\Gamma$ = 6.0 and $\Gamma_0$ = 1.0. 
To consider highly fluctuated cells in their shape, we consider low value of inverse temperature $\beta$ of 0.3. 
Since $\beta$ $\ge$ 0.3 avoids cell fragmentation, this value supports both high fluctuation and stable cells.

By simulating cell dynamics in this model, we examine this model cell for the polarity memory effect. 
Firstly, we decide the simulation strategy on the basis of the theory of the self-propelled disk. 
To the theory,  the motion alignment appears only in the low damping parameter $\gamma$ for the angle of $\bm p_m$, $\theta_m$ . 
Therefore, when the polarity memory effect works, the alignment depends on $\gamma$ in the same manner with the disk and a certain threshold $\gamma_c$ for the alignment is present.
By the reproduction of this dependence, we evaluate the polarity memory effect.

In order to determine $\gamma$, let us derive an effective self-propelled disk from this model.
Since $\gamma$ is independent of intercellular interaction, we consider an isolated cell for simplicity and ignore the intercellular interaction.
In our overdamped simulation based on ${\cal H}$, we can suppose the Langevin equation of cell motion,
\begin{align}
    \frac{d \bm R_m}{dt} 
     = \varepsilon \bm p_m \cdot \sum_{\bm r}\delta_{mm(\bm r)} \left<\frac{\partial \bm e_m(\bm r) }{\partial {\bm r}}\right>_{{\bm R}_m} + \bm f \label{eq:effective_R}
\end{align}
In the derivation of force in rhs, we use the fact that ${\cal H}_{\rm Mot}$ is only the term explicitly depending on $\bm R_m$ in ${\cal H}$. We also {\color{red} interpret the ${\bm R}_m$ derivative as the derivative of the cell boundary coordinate $\bm r$ with fixed ${\bm R}_m$. This is because the force is practically exerted on the cell boundary as defined in the virtual work due to the deformation of cell shape.} 
We divide this force into the shape-average and -fluctuation parts, where $\left< \dots \right >_{{\bm R}_m}$ represents the average over cell shapes with fixed ${\bm R}_m$ and $\bm f$ is deviation representing the shape fluctuation of cells. 
In addition, we reformulate the equation of $\bm p_m$ in Eq.~\eqref{eq:polarity} as an equation of $\theta_m$ similar to the self-propelled disk \cite{Hiraoka:2016}.
For this purpose, we substitute $\bm p_m$ with $(\cos\theta_m, \sin\theta_m)$ and  $d\bm R_m/dt$ with $|d\bm R_m/dt|(\cos\psi_m, \sin\psi_m)$, where $\psi_m$ is the angle of the cell velocity. Then, by multiplying $d{\bm R}_m/dt$ to both sides of Eq.~\eqref{eq:polarity}, we read 
\begin{align}
\frac{d \theta_m}{dt} = - \frac{1}{a\tau_p}\left|\frac{d\bm R_m}{dt}\right| \sin(\theta_m -\psi_m). \label{eq:angle}
\end{align}
The former, Eq.~\eqref{eq:effective_R} is identical to an overdamped active Brownian particle \cite{Martin-Gomez:2018} rather than the self-propelled disk \cite{Hanke:2013, Hiraoka:2016} in the sense of similarity in overdamping feature. In contrast, the latter, Eq.~\eqref{eq:angle} is similar to the self-propelled disk.
In the comparison of Eq.~\eqref{eq:angle} with the corresponding equation in the literature~\cite{Hiraoka:2016}, $|d\bm R/dt|/a\tau_p$ is read as $\gamma$.

In the self-propelled disk, the damping parameter $\gamma$ is a control parameter of the motion alignment \cite{Hanke:2013, Hiraoka:2016}. 
Naively, the dependence on $\gamma$   = $|d\bm R/dt|/a\tau_p$ in our model is evaluable from the $\tau_p$ dependence. 
However, in contrast to self-propelled disk, $\gamma$ also  depends on the cell velocity, $|d\bm R/dt|$, which is not a model parameter.
To consider this effect, we focus on the dependence of the alignment not only on $\tau_p$ and but also $\varepsilon$. 
Here, $\varepsilon$ is one of the parameters which determine $|d\bm R/dt|$ in Eq.~\eqref{eq:effective_R} and the dependence of $|d\bm R/dt|$ on $\varepsilon$ is $|d\bm R/dt|$ $\sim$ $\varepsilon$ for $\varepsilon$ larger than $\bm f$. 

Secondarily, we give the method to identify the motion alignment in our examination.
Refer to the previous work\cite{Vicsek:1995, Hiraoka:2016},
we define the motion alignment by finite values of the order parameter of ${\bm p_m}$,
\begin{align}
     P = \left|\frac{1}{T}\int_{T_0}^{T+T_0} dt \frac{1}{N}\sum_{m=1}^N \bm p_m\right|.
\end{align}
Here, $T_0$ represents the number of mcs for the relaxation to the steady state. $T$ is the number of mcs for the time average. We empirically employ $T_0$ = 10$^6$ mcs and $T$ = 10$^5$ mcs \cite{Matsushita:2017}. 
Here, note that $P$ is not adequate for detection of heterogeneous alignments of motion, including vortexes. Therefore, to avoid the heterogeneous alignments due to the boundary effect of the system, we simply employ the  periodic boundary condition.   

\begin{figure}[t]
    \begin{center}
        \includegraphics[width=0.9\linewidth]{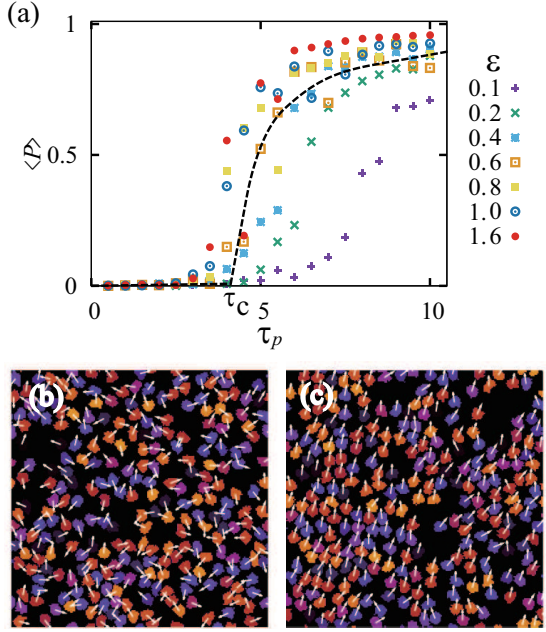}
        \caption{(a) $\left<P\right>$ as a function of time scale ratio $\tau_p$ in Eq.~\eqref{eq:polarity} for various strength of motility $\varepsilon$. The different symbols indicate different values of $\varepsilon$. 
        Snapshots for (b) $\varepsilon$=1.0 and $\tau_p$=$1.0$, and for (c) $\varepsilon$=1.0 and $\tau_p$=$10.0$. Different color region indicates different Potts domains, namely cells. Black region represents the empty region. White arrows represent the direction of polarity.}
        \label{fig:P-Et}
    \end{center}
\end{figure}

To examine the dependence of the alignment on $\tau_p$ and $\varepsilon$, we calculate $P$ as a function of $\tau_p$ for various $\varepsilon$. For easily observing the alignment, we choose relatively high area fraction of cells $\phi$ = $NV/L^2$ $\simeq$ $0.43$. 
The data are plotted in Fig.~\ref{fig:P-Et}(a). $P$ at $\tau_p$ = 0 is almost 0 over all $\varepsilon$. In this case, the cells randomly move and $\bm p_m$ does not align as shown in Fig.~\ref{fig:P-Et}(b).
With increasing $\tau_p$, $P$ is kept to be 0 for small $\tau_p$ and then rapidly increases at a threshold value of $\tau_p$, $\tau_c$ , excepting $\varepsilon$ = 0.1, where the alignment is unstable. 
After this increasing, $P$ takes  high values around unity which indicates the motion alignment shown in Fig.~\ref{fig:P-Et}(c). 
Therefore, even in cells propelled by shape fluctuation, the polarity memory effect works, excepting cases of too small $\varepsilon$'s.

In addition, $\tau_c$ is almost independent of $\varepsilon$ excepting too small value.
This independence seemingly contradicts the expectation from the constant threshold $\gamma_c$ = $|d\bm R/dt|/a\tau_c$ in previous work \cite{Hiraoka:2016}. 
In fact, in the linear response $|d{\bm R_m}/dt|$ $\sim$ $\varepsilon$, the expectation of $\tau_c$ $\sim$ $\varepsilon$ is not shown in Fig.~\ref{fig:P-Et}(a). 
Since the constant threshold $\gamma_c$, which leads to this contradiction, is based on the theory of rigid self-propelled disk, the rigidity of the disk is a candidate origin of this contradiction. 
Therefore, to understand this independence, the shape fluctuation of cells should be considered beyond the theory for the rigid disk.

To explore the origin of this independence, we consider the effect of this shape fluctuation in the motion alignment. 
This effect comes from $\bm f$ in Eq.~\eqref{eq:effective_R} and reduces the polarity memory effect, through $|d\bm R_m/dt|$ and $\psi_m$ in Eq.~\eqref{eq:angle}. 
Intercollision processes reflect this reducing because it shortens the persistent length of cell trajectories, $l_p$, which is defined for an isolated cell differently from the mean free path.  
Since the intercollision processes also reflect the cell density $\phi$ , changes of $\phi$ interferes the effect of $l_p$.
Therefore, by utilizing this interference in the intercollision processes, we can evaluate the shape fluctuation effect. 

\begin{figure}[t]
    \begin{center}
        \includegraphics[width=0.9\linewidth]{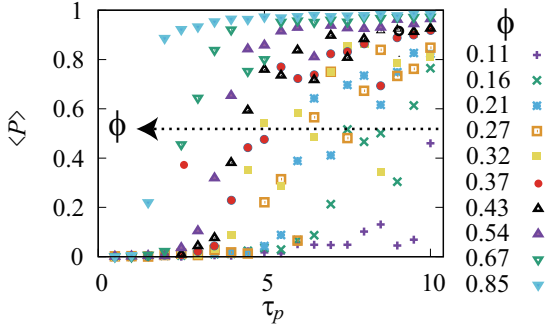}
        \caption{$\left<P\right>$ as a function of time scale ratio $\tau_p$ in Eq.~\eqref{eq:polarity} for various area fraction $\phi$. Different symbols indicate different value of $\phi$. {\color{red}The arrow shows the decreasing of $\tau_c$ with increasing $\phi$.}}
        \label{fig:P-Ft}
    \end{center}
\end{figure}

For this evaluation, we calculate $P$ as a function of $\tau_p$ for various values of $\phi$ with $\varepsilon$ = 1.0 ( $\gg$ $f$ ). 
The result of $P$ is plotted as a function of $\tau_p$ for vairous $\phi$ in Fig.~\ref{fig:P-Ft}. 
$P$ commonly takes 0 for small $\tau_p$ and increases with increasing $\tau_p$. 
The alignment threshold, $\tau_c$, where $P$ rapidly increases, largely decreases with increasing $\phi$ in contrast to the independence on $\varepsilon$ in Fig.~\ref{fig:P-Et}(a). 

Let us consider the mechanism of this strong decreasing $\tau_c$ with increasing $\phi$. 
From discussions so far, we focus on intercollision processes. 
In these processes, since $\bm p_m$ depends on $\bm f$ through $|d{\bm R_m}/dt|$ in Eqs.~\eqref{eq:effective_R} and \eqref{eq:angle}, cells must maintain their direction of polarity against the disorder of shape fluctuation for the stable alignment. 
Therefore, the persistent length, $l_p$, exists as a finite value and determines the stability of alignments. 
This situation is contrast to the self-propelled disks having infinite $l_p$ due to rigid shape\cite{Hiraoka:2016}. 
Hence, for migrating cells, $\tau_c$ is determined not only by $\gamma$ but also by $l_p$. In small shape fluctuation, the memory time of polarity is typically given almost by $t_R$ = $\gamma^{-1}=a\tau_p|d\bm R_m/dt|^{-1}$. This time gives the persistent length,
\begin{align}
    l_p \simeq t_R\left|\frac{d \bm R_m}{d t}\right| = a\tau_p. \label{eq:lp}
\end{align}
$l_p$ must be larger than the intercellular distance $l_d$ for cells to maintain the direction of motion during intercollision processes. 
This gives a necessary condition for the alignment, namely, $l_p$ $>$ $l_d$. 

\begin{figure}[t]
    \begin{center}
        \includegraphics[width=1.0\linewidth]{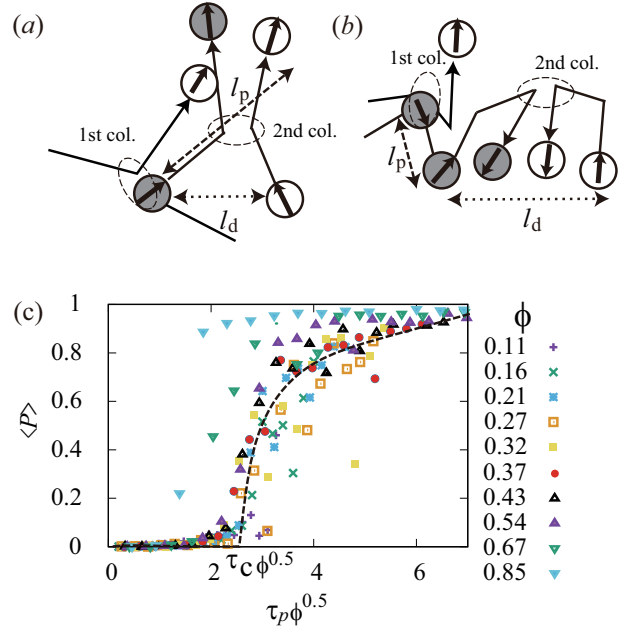}
        \caption{Schematic view of two sequential collision process of cells (grey circle) for (a) $l_p$ $>$ $l_d$ and for (b) $l_p$ $<$ $l_d$. The circles represent a cell and intracircle arrows represent a polarity direction. Solid arrows indicates the cell motion and their straight regions almost correspond to shorter one in $l_d$ and $l_p$. Dashed and dotted lines with arrow heads represent $l_p$ and $l_d$. (c) $\left<P\right>$ as a function of $\tau_p\phi^{1/d}$. Different symbols indicate different value of $\phi$. }
        \label{fig:P-Ft_scaled}
    \end{center}
\end{figure}

To intuitively understand this condition, we discuss the two serial collision processes shown in Figs.~\ref{fig:P-Ft_scaled}(a) and \ref{fig:P-Ft_scaled}(b).
Firstly, we consider the case of $l_p$ $>$ $l_d$.
In this case, as shown in Fig.~\ref{fig:P-Ft_scaled}(a), the direction of motion after the 1st collision is maintained until the 2nd collision, because the 2nd collision typically occurs in the cell movement length of $l_d$. 
Owing to the polarity memory effect, cells align their motion in the common direction through the 1st and 2nd collisions\cite{Hanke:2013,Hiraoka:2016}. 
This aligning repeats in the following collisions and finally results in the motion alignment over all cells. 
Next, we consider the other case of $l_p$ $<$ $l_d$. 
In this case, the polarity relaxes owing to the shape fluctuation in the intercollison periods. 
Therefore, as shown in Fig.~\ref{fig:P-Ft_scaled}(b), even when the motion of cells temporary align in the same direction in the 1st collision, then the aligned direction of motion is lost until the 2nd collision. 
As a result, the motion alignment does not growth over all cells and thereby cannot induces the motion alignment over all cells. 
The marginal case between these cases, $l_p$ $\simeq$ $l_d$, determines $\tau_c$.

On the basis of this marginal condition, we can discuss the independence of $\tau_c$ on $\varepsilon$ in Fig.~\ref{fig:P-Et}(a).
For this discussion, notice that since this system corresponds to the active Brownian particle with high rotational Pe\'{c}let number limit because of the absence of heat bath in Eq.~\eqref{eq:angle} \cite{Martin-Gomez:2018}, 
the phase separation does not appear. 
Therefore, the configuration of cells becomes uniform (See Figs.~\ref{fig:P-Et}(b) and \ref{fig:P-Et}(c)) and thereby $l_d$ is typically $a\phi^{-1/d}$. 
Here, $d$ = 2 is the dimension of space. 
Hence, since $l_d$ = $a\phi^{-1/d}$ and $l_p$ in Eq.~\eqref{eq:lp} are commonly independent of $d{\bm R_m}/dt$, the marginal condition is also independent of $d{\bm R_m}/dt$. 
This is the origin of the independence of $\tau_c$ on $\varepsilon$.  


To confirm this condition, we can use the fact that $l_p/a$ = $\tau_c$ must be equal to $l_d/a$ = $\phi^{-1/d}$. 
In this case, $\tau_c$ decreases with increasing $\phi$ as keeping the value of $l_p$/$l_d$ $=$ $\tau_c\phi^{1/d}$.
In Fig.~\ref{fig:P-Ft_scaled}(c), we replot $P$ in Fig.~\ref{fig:P-Ft} as a function of $\tau_p\phi^{1/d}$. 
In comparison with data in Fig.~\ref{fig:P-Ft}, the data collapse into a single curve, excepting the cases of high area fractions $\phi$ = 0.67 and $\phi$ = 0.85 with crowding effects \cite{Angelini:2011,Bechinger:2016}. 
This result suggests that the crossover of length scales between the  persistent length, $l_p$, and intercellular distance, $l_d$, determines $\tau_c$. 

In conclusion, similarly to the self-propelled disks, cells align their motion by a pure polarity memory effect,  while the condition for alignment has a slight correction due to the shape fluctuation\cite{note:deformable_particle_crossover}.
{\color{red} The polarity effect is a powerful candidate of the alignment mechanism comparably with the shape-velocity coupling effect for the soft deformable particles \cite{Menzel:2012, Yamanaka:2014,Ohta:2017}.}
This implies the biological function of the polarity memory as the driving force of the motion alignment. 
A prominent example is the early development stage of Dicty \cite{Bonner:2009} and therein Dicty exhibits the extention of memory time \cite{Takagi:2008}. 
For utilizing this function in motion alignment, Dicty may strategically extends the memory time of polarity in this early stage.



We thank fruitful discussions with H. Kuwayama, H. Hashimura, S. Yabunaka, K. Hironaka, T. Hiraoka and R. Ishiwata. We also thank the support on the research resource by M. Kikuchi and H. Yoshino.  This work is suppoted by supported by JSPS KAKENHI (Grant Number 19K03770). 


\begin{thebibliography}{10}

\bibitem{Entschladen:2009}
{\em Cell Migration: Signalling and Mechanisms}, ed. F.~Entschladen and K.~S.
  Zanker (S Karger Pub, 2009).

\bibitem{Vicsek:2012}
T.~Vicsek and A.~Zafeiris: Phys. Rep. {\bfseries 517} (2012) 71.

\bibitem{Schaller:2010}
V.~Schaller, C.~Weber, C.~Semmrich, E.~Frey, and A.~R. Bausch: Nature
  {\bfseries 467} (2010) 73.

\bibitem{Seeman:2016}
R.~Seemann, J.-B. Fleury, and C.~C. Maass: Eur. Phys. J. {\bfseries 225} (2016)
  2227.

\bibitem{Brambilla:2013}
M.~Brambilla, E.~Ferrante, M.~Birattari, and M.~Dorigo: Swarm Intell.
  {\bfseries 7} (2013) 1.

\bibitem{Weijer:2009}
C.~J. Weijer: J. Cell Sci. {\bfseries 122} (2015) 3215.

\bibitem{Friedl:2009}
P.~Friedl and D.~Gilmour: Nat. Rev. Mol. Cell Biol. {\bfseries 10} (2009) 445.

\bibitem{Rorth:2009}
P.~R\mbox{\o}rth: Annu. Rev. Cell Dev. Biol. {\bfseries 25} (2009) 407.

\bibitem{Romanczuk:2012}
P.~Romanczuk, M.~BÃ¤r, W.~Ebeling, B.~Lindner, and L.~Schimansky-Geier: Eur.
  Phys. J. {\bfseries 202} (2012) 1.

\bibitem{Marchetti:2013}
M.~C. Marchetti, J.~F. Joanny, S.~Ramaswamy, T.~B. Liverpool, J.~Prost, M.~Rao,
  and R.~A. Simha: Rev. Mod. Phys. {\bfseries 85} (2013) 1143.

\bibitem{Doostmohammadi:2018}
A.~Doostmohammadi, J.~IgnÃ©s-Mullol, J.~M. Yeomans, and F.~Sagu\'{e}s: Nat.
  Comm. {\bfseries 9} (2018) 3246.

\bibitem{Wakita:2015}
J.-I. Wakita, S.~Tsukamoto, K.~Yamamoto, M.~Katori, and Y.~Yamada: J. Phys.
  Soc. Jpn. {\bfseries 84} (2015) 124001.

\bibitem{Kawaguchi:2017}
K.~Kawaguchi, R.~Kageyama, and M.~Sano: Nature {\bfseries 545} (2017) 327.

\bibitem{Wensink:2012}
H.~H. Wensink, J.~Dunkel, S.~Heidenreich, K.~Drescher, R.~E. Goldstein,
  H.~L\"{o}wen, and J.~M. Yeomans: Proc. Natl. Acad. Sci. USA {\bfseries 109}
  (2012) 14308.

\bibitem{Kuwayama:2013}
H.~Kuwayama and S.~Ishida: Sci. Rep. {\bfseries 3} (2013) 2272.

\bibitem{Reynolds:1987}
C.~W. Reynolds: Computer Graphics {\bfseries 21} (1987) 25.

\bibitem{Vicsek:1995}
T.~Vicsek, A.~Czir\'{o}k, E.~Ben-Jacob, I.~Cohen, and O.~Shochet: Phys. Rev.
  Lett. {\bfseries 75} (1995) 1226â�??1229.

\bibitem{Weber:2013}
C.~A. Weber, T.~Hanke, J.~Deseigne, S.~L\'{e}onard, O.~Dauchot, E.~Frey, and
  H.~Chat\'{e}: Phys. Rev. Lett. {\bfseries 110} (2013) 208001.

\bibitem{Lam:2015}
K.-D. N.~T. Lam, M.~Schindler, and O.~Dauchot: New J. Phys. {\bfseries 17}
  (2015) 113056.

\bibitem{Yamanaka:2014}
S.~Yamanaka and T.~Ohta: Phys. Rev. E {\bfseries 90} (2014) 042927.

\bibitem{Ohta:2017}
T.~Ohta: J. Phys. Soc. Jpn. {\bfseries 86} (2017) 072001.

\bibitem{Hanke:2013}
T.~Hanke, C.~A. Weber, and E.~Frey: Phys. Rev. E {\bfseries 88} (2013) 052309.

\bibitem{Hiraoka:2016}
T.~Hiraoka, T.~Shimada, and N.~Ito: Physical Review E {\bfseries 94} (2016)
  062612.

\bibitem{Hiraoka:2017}
T.~Hiraoka, T.~Shimada, and N.~Ito: Journal of Physics: Conference Series
  {\bfseries 921} (2017) 012006.

\bibitem{Martin-Gomez:2018}
A.~Mart�?±\'{n}-Go\'{m}ez, D.~Levis, A.~D�?±\'{a}z-Guilera, and
  I.~Pagonabarraga: Soft Matter {\bfseries 14} (2018) 2610.

\bibitem{Petrie:2009}
R.~J. Petrie, A.~D. Doyle, and K.~M. Yamada: Nature Rev. Mol. Cell Biol.
  {\bfseries 10} (2009) 538.

\bibitem{Trepat:2018}
X.~Trepat and E.~Sahai: Nat. Phys. {\bfseries 14} (2018) 671.

\bibitem{Takagi:2008}
H.~Takagi, M.~J. Sato, T.~Yanagida, and M.~Ueda: PLOS One {\bfseries 3} (2008)
  e2648.

\bibitem{Li:2011}
L.~Li, E.~C. Cox, and H.~Flyvbjerg: Phys. Biol. {\bfseries 8} (2011) 046006.

\bibitem{Szabo:2006}
B.~Szab\'{o}, G.~J. Szollosi, B.~Gonci, Z.~Juranyi, D.~Selmeczi, and T.~Vicsek:
  Phys.~Rev.~E {\bfseries 74} (2006) 061908.

\bibitem{Maree:2006}
A.~F.~M. Mar\'{e}e, A.~Jilkine, A.~Dawes, Ver\^{o}nica, A.~Grieneisen, and
  L.~Edelstein-Keshet: Bull. Math. Biol. {\bfseries 68} (2006) 1169â�??1211.

\bibitem{Szabo:2016}
A.~Szab\'{o}, M.~Melchionda, G.~Nastasi, M.~L. Woods, S.~Campo, R.~Perris, and
  R.~Mayor: J. Cell Biol. {\bfseries 215} (2016) 543.

\bibitem{Mehes:2013}
E.~M\'{e}hes and T.~Vicsek: Comput.. Adapt. Syst. Mod. {\bfseries 1} (2013) 4.

\bibitem{Haeger:2015}
A.~Haeger, K.~Wolf, M.~M. Zegers, and P.~Friedl: Trends Cell Biol. {\bfseries
  25} (2015) 556.

\bibitem{Bagorda:2008}
A.~Bagorda and C.~A. Parent: J. Cell Sci. {\bfseries 121} (2008) 2621.

\bibitem{Akiyama:2017}
M.~Akiyama, T.~Sushida, S.~Ishida, and H.~Haga: Develop. Growth Differ.
  {\bfseries 59} (2017) 471.

\bibitem{Kuwayama:1993}
H.~Kuwayama, S.~Ishida, and P.~J. M.~V. Haastert: J. Cell Biol. {\bfseries 123}
  (1993) 1453.

\bibitem{Ilina:2009}
O.~Ilina and P.~Friedl: J. Cell Sci. {\bfseries 122} (2009) 3203.

\bibitem{Matsushita:2018}
K.~Matsushita: Phys. Rev. E {\bfseries 97} (2018) 042413.

\bibitem{Merks:2006}
R.~M. Merks, S.~V. Brodsky, M.~S. Goligorksy, S.~A. Newman, and J.~A. Glazier:
  Develop. Biol. {\bfseries 289} (2006) 44.

\bibitem{Starruss:2007}
J.~S.~T. Bley, L.~S\mbox{\o}gaard-Andersen, and A.~Deutsch: J. Stat. Phys.
  {\bfseries 128,} (2007) 269.

\bibitem{Hirashima:2017}
T.~Hirashima, E.~G. Rens, and R.~M.~H. Merks: Development, growth \&
  differentiation {\bfseries 59} (2017) 329.

\bibitem{Mayor:2010}
R.~Mayor and C.~Carmona-Fontaine: Trends Cell Biol. {\bfseries 20} (2010) 319.

\bibitem{Schnyder:2017}
S.~K. Schnyder, J.~J. Molina, Y.~Tanaka, and R.~Yamamoto: Sci Rep. {\bfseries
  7} (2017) 5163.

\bibitem{Fujimori:2019}
T.~Fujimori, A.~Nakajima, N.~Shimada, and S.~Sawai: Proc. {\bfseries 116}
  (2019) 4291.

\bibitem{Hiraiwa:2019}
T.~Hiraiwa: Phys. Rev. E {\bfseries 99} (2019) 012614.

\bibitem{Blanchard:2010}
G.~B. Blanchard, S.~Murugesu, R.~J. Adams, A.~Martinez-Arias, and
  N.~Gorfinkiel: Development {\bfseries 137} (2010) 2743.

\bibitem{Munoz:2013}
A.~Mu\mbox{\~{n}}oz, D.~A. Fletcher, and O.~D.Weiner: Trends in Cell Biology
  {\bfseries 23} (2013) 47.

\bibitem{Bignold:1987}
L.~P. Bignold: Experientia {\bfseries 43} (1987) 860.

\bibitem{Haastert:2011}
P.~J. M.~V. Haastert: Sci. Signal. {\bfseries 4} (2011) pe6.

\bibitem{Fackler:2008}
O.~T. Fackler and R.~Grosse: J. Cell. Biol. {\bfseries 181} (2008) 879.

\bibitem{Charras:2008}
G.~Charras and E.~Paluch: Nature Rev. Mol. Cell Biol. {\bfseries 9} (2008)
  730â�??736.

\bibitem{Anderson:2007}
A.~R.~A. Anderson, M.~A.~J. Chaplain, and K.~A. Rejniak: {\em Single-Cell-Based
  Models in Biology and Medicine} (Birkhauser Verlag AG, Basel, 2007).

\bibitem{Graner:1992}
F.~Graner and J.~A. Glazier: Phys.~Rev.~Lett. {\bfseries 69} (1992) 2013.

\bibitem{Graner:1993}
F.~Graner: J. Theor. Biol. {\bfseries 164} (1993) 455.

\bibitem{Kabla:2012}
A.~J. Kabla: J. R. Soc. Interface {\bfseries 9} (2012) 3268.

\bibitem{Glazier:1993}
J.~A. Glazier and F.~Graner: Phys. Rev. E {\bfseries 47} (1993) 2128.

\bibitem{Matsushita:2017}
K.~Matsushita: Phys. Rev. E. {\bfseries 95} (2017) 032415.

\bibitem{Angelini:2011}
T.~E. Angelini, E.~Hannezo, X.~Trepat, M.~Marquez, J.~J. Fredberg, and D.~A.
  Weitz: Proc {\bfseries 108} (2011) 4714.

\bibitem{Bechinger:2016}
C.~Bechinger, R.~D. Leonardo, H.~L\"{o}wen, C.~Reichhardt, G.~Volpe, and
  G.~Volpe: Rev. Mod. Phys {\bfseries 88} (2016) 045006.

\bibitem{note:deformable_particle_crossover}
The connection from cells to the self-propelled disks with the continuous
  change in the shape rigidity is unknown. The simple rigid-limit of these
  model cells does not correspond to the self-propelled disks because the rigid
  cells lose propulsion based on the shape deformation. Therefore, the solution
  of this problem is outside the scope of our model and needs other suitable
  model.

\bibitem{Menzel:2012}
A.~M. Menzel and T.~Ohta: Europhys. Lett. {\bfseries 99} (2012) 58001.

\bibitem{Bonner:2009}
J.~T. Bonner: {\em The Social Amoebae: The Biology of Cellular Slime Molds}
  (Princeton University Press, Princeton, 2009).

\end{thebibliography}

\end{document}